\documentclass[aps,showpacs,superscriptaddress,preprint]{revtex4}%
\usepackage{amsfonts}
\usepackage{amsmath}
\usepackage{amssymb}
\usepackage{graphicx}%
\providecommand{\U}[1]{\protect\rule{.1in}{.1in}}

\begin{document}
\title{Elasticity behavior, phonon spectra, and the pressure-temperature phase
diagram of HfTi alloy: A density-functional theory study }
\author{Yong Lu}
\affiliation{LCP, Institute of Applied Physics and Computational Mathematics, Beijing
100088, People's Republic of China}
\author{Ping Zhang}
\thanks{Author to whom correspondence should be addressed. E-mail: zhang\_ping@iapcm.ac.cn}
\affiliation{LCP, Institute of Applied Physics and Computational Mathematics, Beijing
100088, People's Republic of China}


\begin{abstract}
The pressure-induced phase transition, elasticity behavior, thermodynamic
properties, and $P\mathtt{-}T$ phase diagram of $\alpha$, $\omega$, and
$\beta$ equiatomic HfTi alloy are investigated using first-principles
density-functional theory (DFT). The simulated pressure-induced phase
transition of the alloy follows the sequence of $\alpha\mathtt{\rightarrow
}\omega\mathtt{\rightarrow}\beta$, in agreement with the experimental results
of Hf and Ti metals. Our calculated elastic constants show that the $\alpha$
and $\omega$ phases are mechanically stable at ambient pressure, while the
$\beta$ phase is unstable, where a critical pressure of 18.5 GPa is predicted
for its mechanical stability. All the elastic constants, bulk modulus, and
shear modulus increase upon compression for the three phases of HfTi. The
ductility of the alloy is shown to be well improved with respect to pure Hf
and Ti metals. The Mulliken charge population analysis illustrates that the
increase of the $d$-band occupancy will stabilize the $\beta$ phase under
pressure. The phonon spectra and phonon density of states are studied using
the supercell approach for the three phases, and the stable nature of $\alpha$
and $\omega$ phases at ambient pressure are observed, while the $\beta$ phase
is only stable along the [110] direction. With the Gibbs free energy
calculated from DFT-parametrized Debye model as a function of temperature and
pressure, the phase transformation boundaries of the $\alpha$, $\omega$, and
$\beta$ phases of HfTi are identified.

\end{abstract}
\maketitle

\section{INTRODUCTION}

Group-IV transition metals and alloys have attracted great scientific and
technological interests since their particular applications in the aerospace,
atomic energy industry, nuclear reactor, and chemical industry. The three
Group-IV metals that occur naturally are titanium (Ti), zirconium (Zr) and
hafnium (Hf). Titanium is recognized for its high strength-to-weight ratio
with low density, meanwhile, the corrosion resistance, the heat stability, and
the ductility are of benefit. The foremost use of hafnium and zirconium has
been in nuclear reactors due to their corrosion resistance. Hafnium has a high
thermal neutron-capture cross-section while zirconium possesses a rather low
one, therefore, they can be used as control rod and cladding of fuel rods in
nuclear reactors respectively \cite{Schemel,Hedrick,Spink}. Properly, hafnium
and zirconium are used in nickel-based super alloys to improve their
mechanical properties \cite{Donachie}.

The titanium, hafnium and zirconium are complete solid solution between each
other. The appropriate solution of these metals will help to improve the
mechanical or thermal properties. Scientifically, most interests are attracted
in their narrow $d$-band in the midst of a broad $sp$-band, where an increase
in $d$-electron population by transfer from the $s$ band is the driving force
behind the structural and electronic transitions \cite{Duthie,Skiver}. The
pressure-induced phase transformation sequence has received extensive
experimental as well as theoretical attention \cite{Hixson, Vohra1990, Xia,
Xia2}. At room temperature and under compression, Hf undergoes a
crystallographic phase transition from hcp ($\alpha$ phase) to the hexagonal
structure ($\omega$ phase) at about 38$\pm$8 GPa \cite{Xia}. Upon further
compression, Hf has been observed to transform into the bcc structure ($\beta$
phase) at 71$\pm$1 GPa \cite{Xia}. For Ti, the measured phase transition
sequence at room temperature is $\alpha\rightarrow\omega\rightarrow
\gamma\rightarrow\delta$ \cite{Xia2, Vohra, Akahama,Errandonea}, and the
$\beta$ phase has not yet been observed up to 216 GPa \cite{Akahama}. However,
the recent theoretical investigation \cite{Mei} found that the $\delta$ phase
is not stable under hydrostatic compression, and the $\delta$ phase should be
replaced by $\beta$ phase at zero Kelvin. The absence of the high-pressure
$\beta$ phase for Ti in experiments was attributed to the possible
nonhydrostatic stress which distorts the $\beta$ phasse \cite{Verma}.

The equilibrium phases at ambient pressure of the Hf-Ti system have been
tabulated by several studies \cite{Imgram, Murray, Bittermann, Hayes,
Tylkina}, containing the liquid, and the $\beta$ and $\alpha$ phases of the
solid. The $\alpha\rightarrow\beta$ phase transition for the equiatomic HfTi
alloy has been measured to be $\sim$1200 K at ambient pressure. However, up to
now, the effects of pressure on the phase transition of HfTi alloy have not
been reported yet. The stability of the $\alpha$, $\omega$, and $\beta$ phases
of HfTi alloy also need for testing to support their practical application.
Thus, in the present study, our main task is to investigate the
pressure-induced phase transition, the elasticity behaviors upon pressure, and
the thermodynamic properties of equiatomic HfTi alloy. As well, the
$P\mathtt{-}T$ phase diagram are also predicted. The rest of the paper is
organized as follows. The theory of Helmholtz energy calculation in the
quasiharmonic approximation and computational details of first-principles are
briefly introduced in Section II. The calculation results are presented and
discussed in Section III. Finally, we give a summary of this work in Section IV.

\section{Theory and Calculation Methods}

The Helmholtz free energy \emph{F} can be approximated as
\begin{equation}
F(V,T)=E(V)+F_{vib}(V,T)+F_{ele}(V,T),\label{eq1}%
\end{equation}
where $E(V)$ stands for the ground-state cold energy, $F_{vib}(V,T)$ is the
vibrational energy of the lattice ions at a given unit cell volume $V$, and
$F_{ele}$ is the thermal electronic contribution to the free energy. Under
quasihamonic approximation \cite{Siegel}, the $F_{vib}(V,T)$ can be evaluated
from phonon density of states (DOS) $g(\omega)$ by
\begin{equation}
F_{ph}(V,T)=k_{B}T\int_{0}^{\infty}g(\omega)\ln\left[  2\sinh\left(
\frac{\hbar\omega}{2k_{B}T}\right)  \right]  d\omega,\label{pho}%
\end{equation}
where $\omega$=$\omega(V)$ depends on volume and thus Equation (\ref{pho})
contains some effect of anharmonics, and $g(\omega)$ is the phonon DOS which
should be positive. So, this formula is not suitable for dynamically unstable
phases. Instead, the Debye model can be employed to estimate the vibrational
energy for phases with imaginary phonon frequencies by
\begin{equation}
F_{vib}(V,T)=\frac{9}{8}k_{B}\Theta_{D}+k_{B}T\left[  3\ln(1-exp(-\frac
{\Theta_{D}}{T})-D(\frac{\Theta_{D}}{T})\right]  ,\label{debye}%
\end{equation}
where $\frac{9}{8}k_{B}\Theta_{D}$ is the zero-point energy due to lattice ion
vibration at 0 K, and $D(\frac{\Theta_{D}}{T})$ is the Debye function given by
$D(\frac{\Theta_{D}}{T})=\frac{3}{x^{3}}\int_{0}^{\Theta/T}x^{3}/(e^{x}-1)dx$
as introduced in Ref. \cite{Blanco} explicitly. $F_{ele}$ in Equation
(\ref{eq1}) can be obtained from the energy and entropy contributions,
\emph{i.e.}, $E_{ele}-TS_{ele}$. The electronic entropy $S_{ele}$ is of the
form
\begin{equation}
S_{ele}(V,T)=-k_{B}\int{n(\varepsilon,V)[f\ln{f}+(1-f)\ln{(1-f)}]d\varepsilon
},
\end{equation}
where \emph{n}($\varepsilon$) is electronic DOS, and ${{f}}$ is the
Fermi-Dirac distribution. The energy \emph{E$_{ele}$} due to the electron
excitations takes the following form
\begin{equation}
E_{ele}(V,T)=\int{n(\varepsilon,V)f\varepsilon d\varepsilon}-\int
^{\varepsilon_{F}}{n(\varepsilon,V)\varepsilon d\varepsilon},
\end{equation}
where $\varepsilon_{F}$ is the Fermi energy.

The DFT calculations are carried out using the Vienna \textit{ab initio}
simulations package (VASP) \cite{G.Kresse1,G.Kresse2} with the
projector-augmented-wave (PAW) potential methods \cite{PAW}. The exchange and
correlation effects are described by generalized gradient approximation (GGA)
in the Perdew-Burke-Ernzerhof (PBE) form \cite{PBE} and the plane-wave basis
set is limited by the cutoff energy of 500 eV. The integration over the
Brillouin Zone (BZ) is done on 18$\times$18$\times$16, 16$\times$16$\times$9,
and 18$\times$18$\times$18 $k$-point meshes generated by the Monkhorst-Pack
\cite{Monkhorst} method for $\alpha$ (two-atom cell), $\omega$ (six-atom
1$\times$1$\times$2 supercell), and $\beta$ (two atoms cell) phases,
respectively. Full geometry optimization at each volume is considered to be
completed when the energy convergence and Hellmann-Feynman forces become less
than 1.0$\times$10$^{-5}$ eV/atom and 0.01 eV/{\AA }, respectively.

\section{Results and discussions}

\subsection{Ground state properties}

\begin{figure}[ptb]
\includegraphics[width=1.0\textwidth]{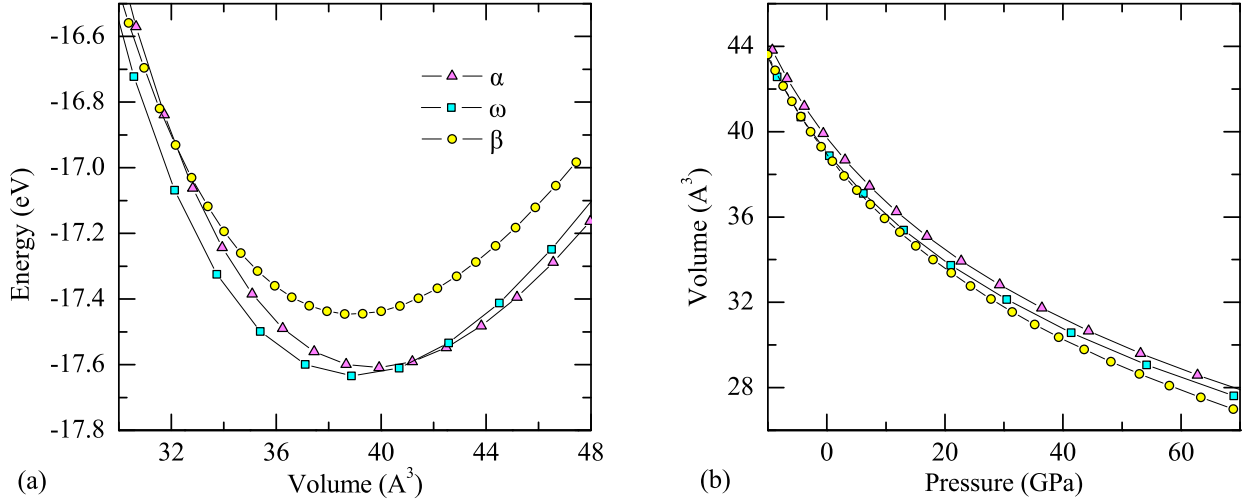}\caption{Calculated (a)
ground-state energy as a function of volume per formula unit cell, and (b)
volume as a function of pressure for $\alpha$, $\omega$, and $\beta$ HfTi
alloy. }%
\label{fig:fig1}%
\end{figure}

The total energies of the $\alpha$, $\omega$, and $\beta$ phases at different
volumes are calculated and shown in Fig. 1(a). Obviously, the $\omega$ phase
is estimated to be the most stable structure, and the $\beta$ phase most
unstable. The energy-volume curve of $\alpha$ phase intersects with that of
$\beta$ and $\omega$ phases at different volumes respectively. The volumes of
the three phases as a function of pressure are depicted in Fig. 1(b). Clearly,
the volume of $\alpha$ phase is always the largest at the considered pressure
range, while the volume-pressure curves of $\omega$ and $\beta$ phases have an
intersection at around $-$4.4 GPa. In order to obtain the theoretical
equilibrium lattice parameter ($a$), bulk modulus ($B$), and pressure
derivative of bulk modulus ($B^{\prime}$) of the three phases, we fit their
energy-volume data to the third-order Birch-Murnaghan equation of states (EOS)
\cite{Brich}. The fitting results are tabulated in Table I, together with the
experimental values \cite{Vohra, Kittel, Donohue, Sikka, Tonkov, Ostanin,
Chang, Rudy, Fisher, JZZhang} for comparison. The corresponding data of pure
Hf and Ti metals are also calculated and listed in Table I. Evidently, for
$\alpha$ phase, one can find excellent coincidence between our calculated
values and the corresponding experimental results of equilibrium lattice
parameters $a$ and $c/a$ ratio for both the alloy and the pure metals. The
calculated bulk modulus $B$ of 110.6 GPa for $\alpha$-HfTi alloy lies within
the range between $\alpha$-Hf (108.2 GPa) and $\alpha$-Ti (113.0 GPa). For the
$\omega$ phase, our calculated equilibrium crystal constants of Hf and Ti are
both in agreement with the corresponding experimental results within 1\%
error, and a value of 4.774 \AA $~$for their alloy is obtained. As for the
high temperature $\beta$ phase, our calculated lattice parameter of Hf is
lower than the experimental value by 1.9 \%, the reason of which can be
attributed to the lattice thermal expansion with temperature. Similar to other
two phases, the equilibrium lattice parameter of $\beta$-HfTi lies within the
corresponding values of Hf and Ti.

\begin{table}[ptb]
\caption{Calculated lattice constants ($a$ and $c/a$), bulk modules ($B$),
pressure derivative of bulk modulus ($B^{\prime}$), and elastic constants of
$\alpha$-, $\omega$-, and $\beta$-phase HfTi, Hf, and Ti at ambient pressure.
For comparison, experimental results are also listed.}%
\begin{ruledtabular}
\begin{tabular}{cccccccccccccccc}
&Phase&Method&$a$&$c/a$&$B$&$B'$&$C_{11}$&$C_{12}$&$C_{13}$&$C_{33}$&$C_{44}$\\
&&&(\AA)&&(GPa)&&(GPa)&(GPa)&(GPa)&(GPa)&(GPa)\\
\hline
HfTi&$\alpha$&This study&3.122&1.555&110.6&3.61&194.0&66.1&75.8&193.0&45.1\\
&&Expt.&3.08&1.571$^{a,b}$&&&\\
&$\omega$&This study&4.774&0.619&117.8&3.07&194.9&81.8&58.3&245.2&49.9\\
&$\beta$&This study&3.402&&107.5&3.26&99.5&120.8&&&38.9\\
Hf&$\alpha$&This study&3.202&1.581&108.2&3.37&194.0&59.0&68.8&196.2&52.7\\
&&Expt.&3.190&1.583$^{c,d}$&&&190.1&74.5&65.5&204.4&60.0$^{e}$\\
&$\omega$&This study&4.989&0.621&109.3&3.45&200.3&76.4&47.6&240.4&49.1\\
&&Expt.&4.943&0.617$^{f}$&&&\\
&$\beta$&This study&3.545&&95.5&3.58&72.5&115.1&&&51.8\\
&&Expt.&3.615$^{g}$&\\
Ti&$\alpha$&This study&2.939&1.583&113.0&3.424&194.4&63.6&77.1&188.8&42.8\\
&&Expt.&2.957&1.585&102.0&3.9$^{h}$\\
&&&2.95&1.586&114.0(3)&4.0$^{i}$\\
&&&2.951&1.587&&&162.4&92.0&69.0&180.7&46.7$^{j}$\\
&&&&&&&176&86.9&68.3&191&50.8$^{e}$\\
&$\omega$&This study&4.580&0.618&111.5&3.51&195.9&84.6&55.0&243.6&53.3\\
&&Expt.&4.598&0.614&142.0$^{h}$&&&\\
&$\beta$&This study&3.25&&105.7&3.33&93.6&115.9&&&39.8\\
\end{tabular} \label{a}
$^{a}$ Ref. \cite{Chang}, $^{b}$ Ref. \cite{Rudy}, $^{c}$ Ref.
\cite{Kittel}, $^{d}$ Ref. \cite{Donohue}, $^{e}$, Ref.
\cite{Fisher}, $^{f}$ Ref. \cite{Sikka}, $^{g}$ Ref. \cite{Tonkov},
$^{h}$ Ref. \cite{Vohra}, $^{i}$ Ref. \cite{JZZhang}, $^{j}$ Ref.
\cite{Ostanin}.
\end{ruledtabular}
\end{table}

\subsection{Pressure induced phase transition}

\begin{figure}[ptb]
\includegraphics[width=0.8\textwidth]{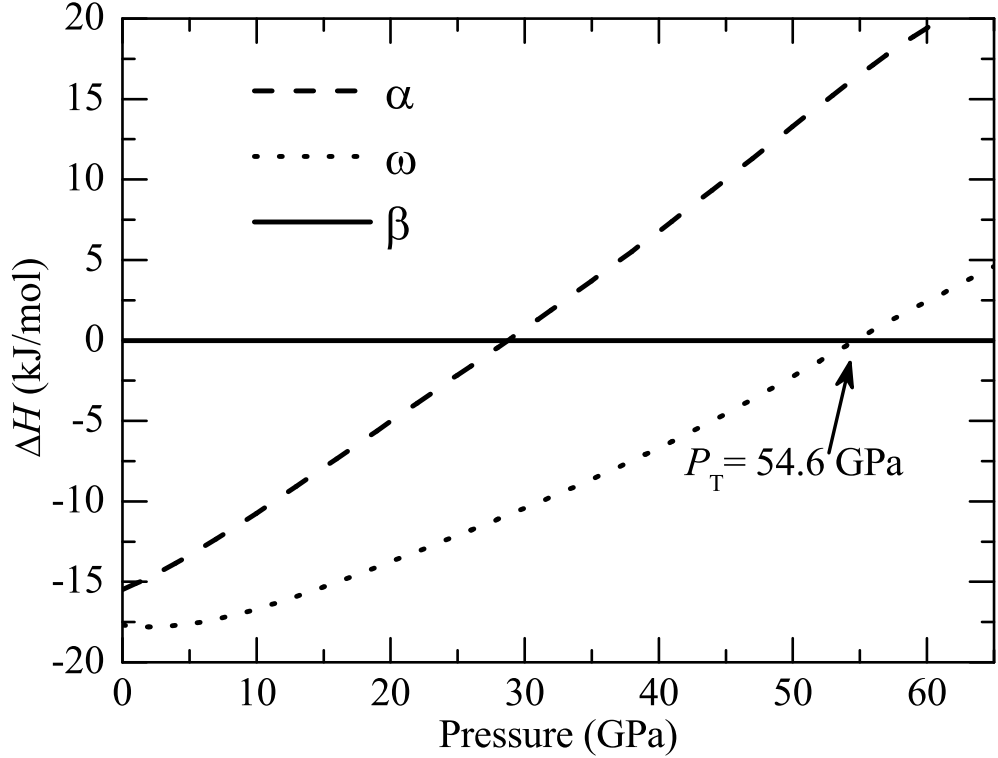}\caption{Calculated enthalpy
differences of $\alpha$ and $\omega$ phases with respect to $\beta$ phase as a
function of pressure. }%
\label{fig:fig2}%
\end{figure}

\begin{figure}[ptb]
\includegraphics[width=0.8\textwidth]{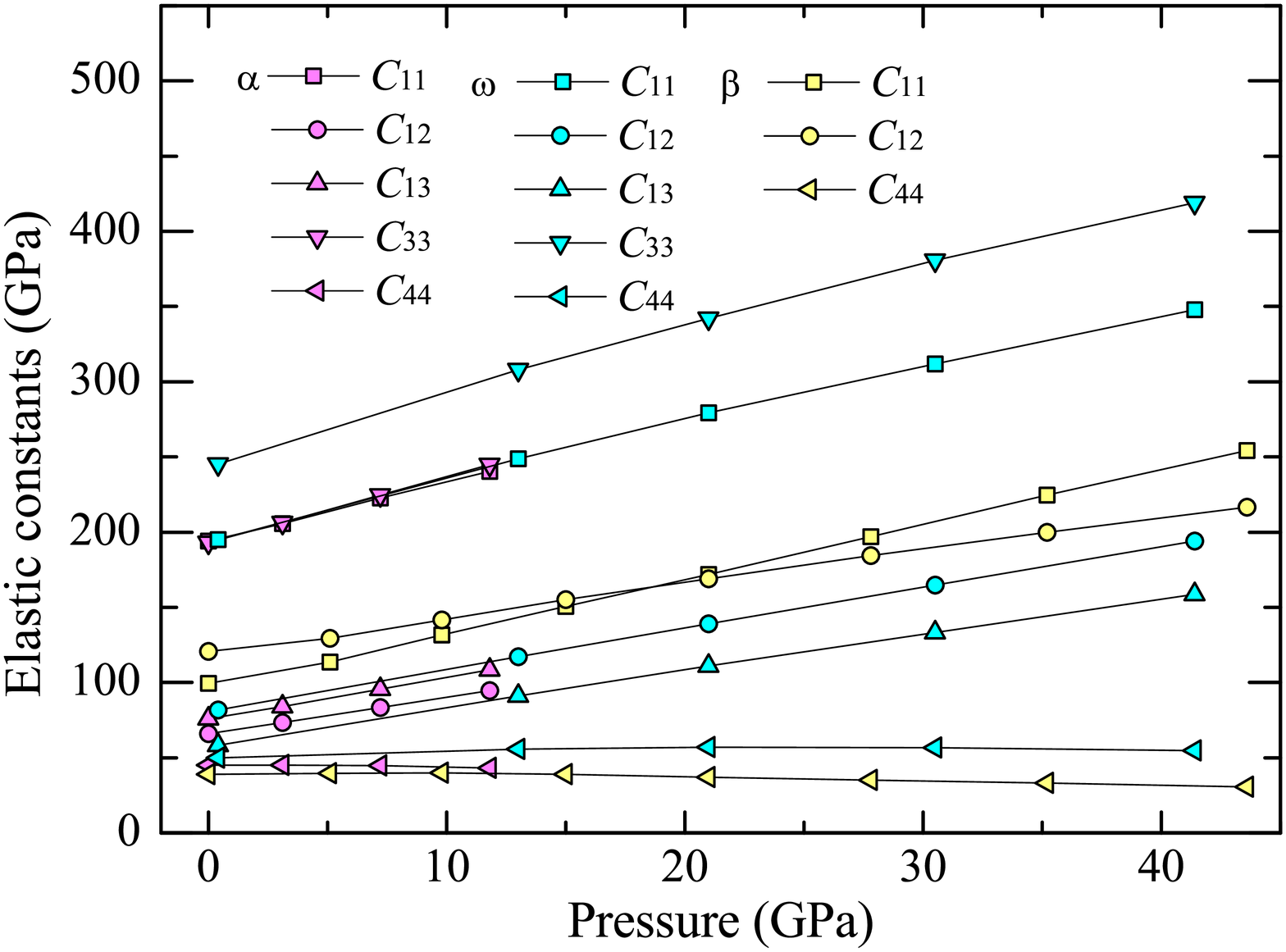}\caption{Calculated elastic
constants as a function of pressure for $\alpha$, $\omega$, and $\beta$ HfTi
alloy. }%
\label{fig:fig3}%
\end{figure}

As shown in Fig. 1(a), the intersection of energy-volume curves
between $\alpha$ phase and other two phases indicate that the phase
transformation will occur between them at specific pressures.
Theoretically, the transition pressure between $\alpha$ and $\omega$
phases can be obtained from the common tangent of their
energy-volume curves. However, it is difficult to determine the slop
accurately. Optionally, we can obtain the transition pressure by
comparing their Gibbs free energy. At 0 K, the Gibbs free energy is
equal to the enthalpy $H\mathtt{=}E\mathtt{+}PV$. In Fig. 2 we plot
the enthalpies of the $\alpha$ and $\omega$ phases with respect to
the $\beta$ phase as a function of pressure. Clearly, at ambient
pressure and above the $\omega$ phase is more stable than the
$\alpha$ one, and there is no crossing between them. Indeed, in our
calculation, the crossing between the $\omega$ and $\alpha$ enthalpy
curves lies at the pressure of $-$4.3 GPa. This result is consistent
with the theoretical result of Ti \cite{Mei,Kutepov}, however, it is
inconsistent with the transition sequence of Hf both experimentally
and theoretically \cite{Ahuja, Jomard, Hao, Xia}. To further
investigate the phase stability, in Table II we list the transition
pressures for HfTi alloy as well as for pure Hf and Ti metals from
both experiments and theoretical calculations. For metal Hf, at
ambient pressure both the experimental \cite{Xia} and theoretical
studies \cite{Ahuja, Jomard, Hao} indicate that the most stable
phase is the $\alpha$ phase. The measured transition pressure of
$\alpha\mathtt{\rightarrow}\omega$ is 38$\pm$8 GPa \cite{Xia}, and
the theoretical results vary from 13.9 GPa \cite{Ahuja} to 43.5 GPa
\cite{Jomard} and 44.5 GPa \cite{Hao}. For metal Ti, there exist
debates in theoretical studies \cite{Mei,Kutepov, Jona}, although
experiments \cite{Vohra, Xia2, Akahama, Errandonea} have reported
the most stable phase to be $\alpha$ phase, as shown in Table II. We
find that the theoretical DFT-PBE studies \cite{Mei,Kutepov}, giving
the negative $\alpha\mathtt{\rightarrow}\omega$ transition
pressures, are performed at 0 K. While, by considering the
temperature, the DFT-PBE study \cite{Mei2} explicitly show that the
$\alpha\mathtt{\rightarrow}\omega$ transition occurs at $\sim$1.8
GPa at room temperature. Thus, the disagreement between some
theoretical studies and measured values for Ti metal mainly
originates from the effect of temperature. For HfTi alloy, the phase
transition of $\omega\mathtt{\rightarrow}\beta$ occurs at 54.6 GPa.
This value coincides with the theoretical results of 30.7-66.2 GPa
\cite{Ahuja, Jomard, Hao} for Hf metal, and is somewhat lower than
the experimental data of 71 GPa \cite{Xia}. Also, the temperature
effect on phase transition pressure cannot be ignored. Thus, here we
will consider the effect of temperature on the transition sequence
of HfTi alloy by employing the same scheme as in our previous
studies of Zr metal \cite{Wang} and TiZr alloy \cite{Wang2}.

\begin{table}[ptb]
\caption{Calculated transition pressure of HfTi. For comparison, other
theoretical results and experimental data for Hf and Ti are listed.}%
\begin{ruledtabular}
\begin{tabular}{cccccccccccccccc}
&Phase transition&Theory&Expt.\\
&&(GPa)&(GPa)\\
\hline\\
HfTi&$\alpha \rightarrow \omega$&-4.3$^{a}$&\\
&$\omega \rightarrow \beta$&54.6$^{a}$&\\
Hf&$\alpha \rightarrow \omega$&13.9$^{b}$, 43.5$^{c}$, 44.5$^{d}$&38$\pm$8$^{e}$\\
&$\omega \rightarrow \beta$&30.7$^{b}$, 62.6$^{c}$, 66.2$^{d}$&71$^{e}$\\
Ti&$\alpha \rightarrow \omega$&-3.7$^{f}$, -3.0$^{g}$, 52$^{h}$&2-11.9$^{i}$\\
\end{tabular} \label{a}
$^{a}$ This study, $^{b}$ Ref. \cite{Ahuja}, $^{c}$ Ref.
\cite{Jomard}, $^{d}$ Ref. \cite{Hao}, $^{e}$ Ref. \cite{Xia},
$^{f}$ Ref. \cite{Mei}, $^{g}$ Ref. \cite{Kutepov}, $^{h}$ Ref.
\cite{Jona}, $^{i}$ Refs. \cite{Vohra, Xia2, Akahama, Errandonea}
\end{ruledtabular}
\end{table}

\subsection{Elasticity behavior at high pressure}

\begin{figure}[ptb]
\includegraphics[width=0.8\textwidth]{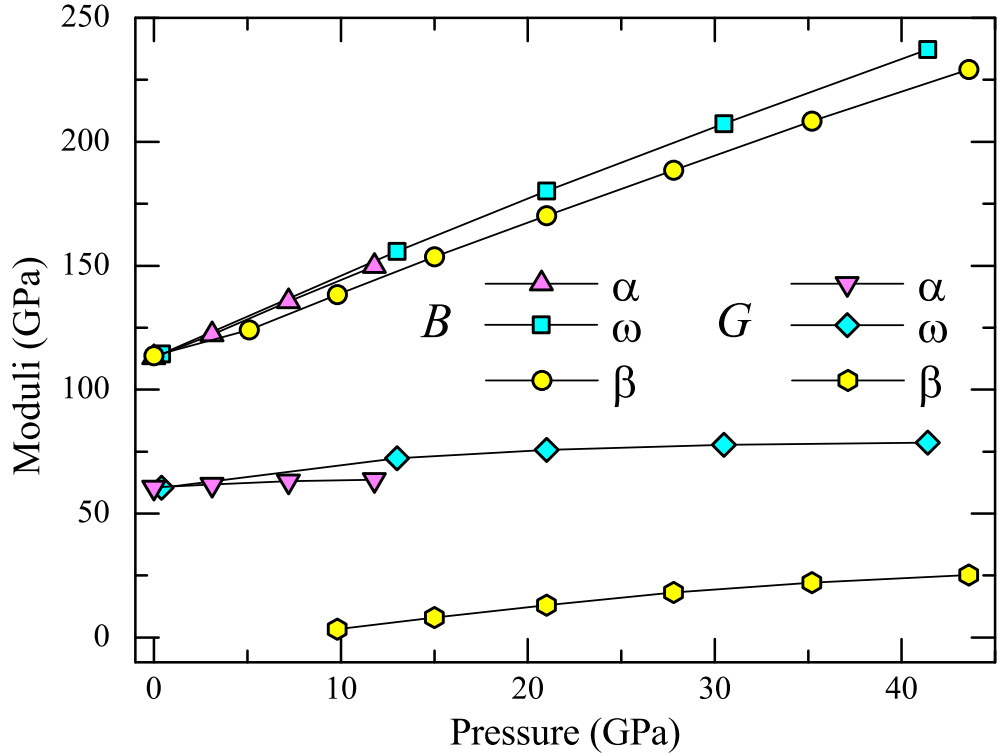}\caption{Calculated bulk modulus
($B$) and shear modulus ($G$) as a function of pressure for $\alpha$, $\omega
$, and $\beta$ HfTi alloy.}%
\label{fig:fig4}%
\end{figure}

\begin{figure}[ptb]
\includegraphics[width=0.8\textwidth]{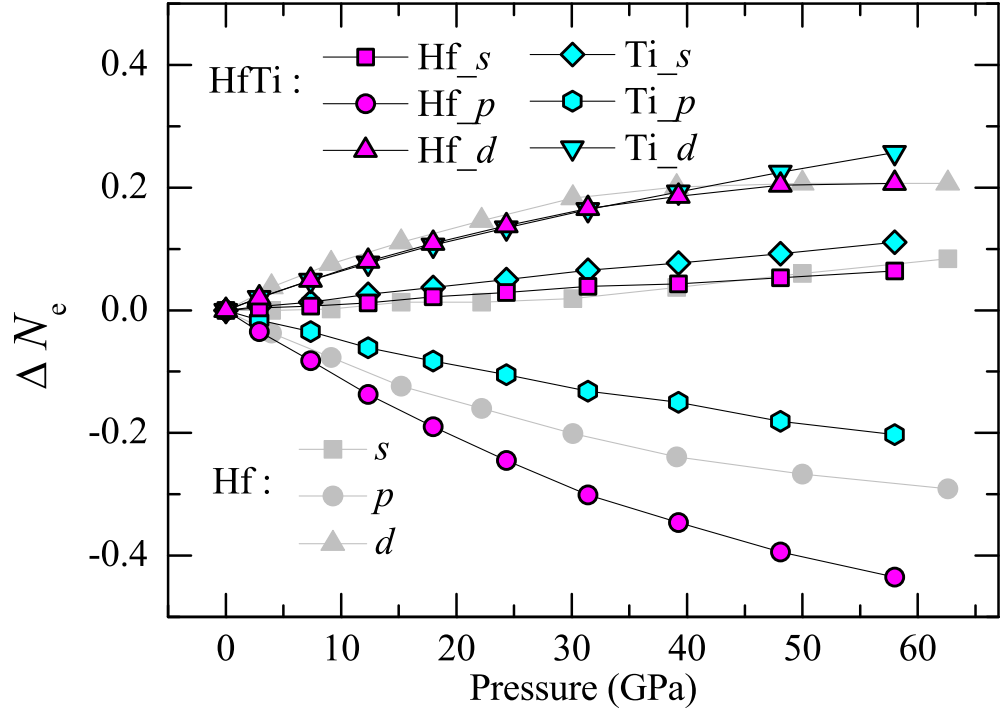}\caption{The number of electrons
on $s$, $p$, and $d$ orbitals for $\beta$ phase HfTi alloy and Hf metal as
functions of pressure, with respect to to their values at ambient pressure.}%
\label{fig:fig5}%
\end{figure}

\begin{figure}[ptb]
\includegraphics[width=1.0\textwidth]{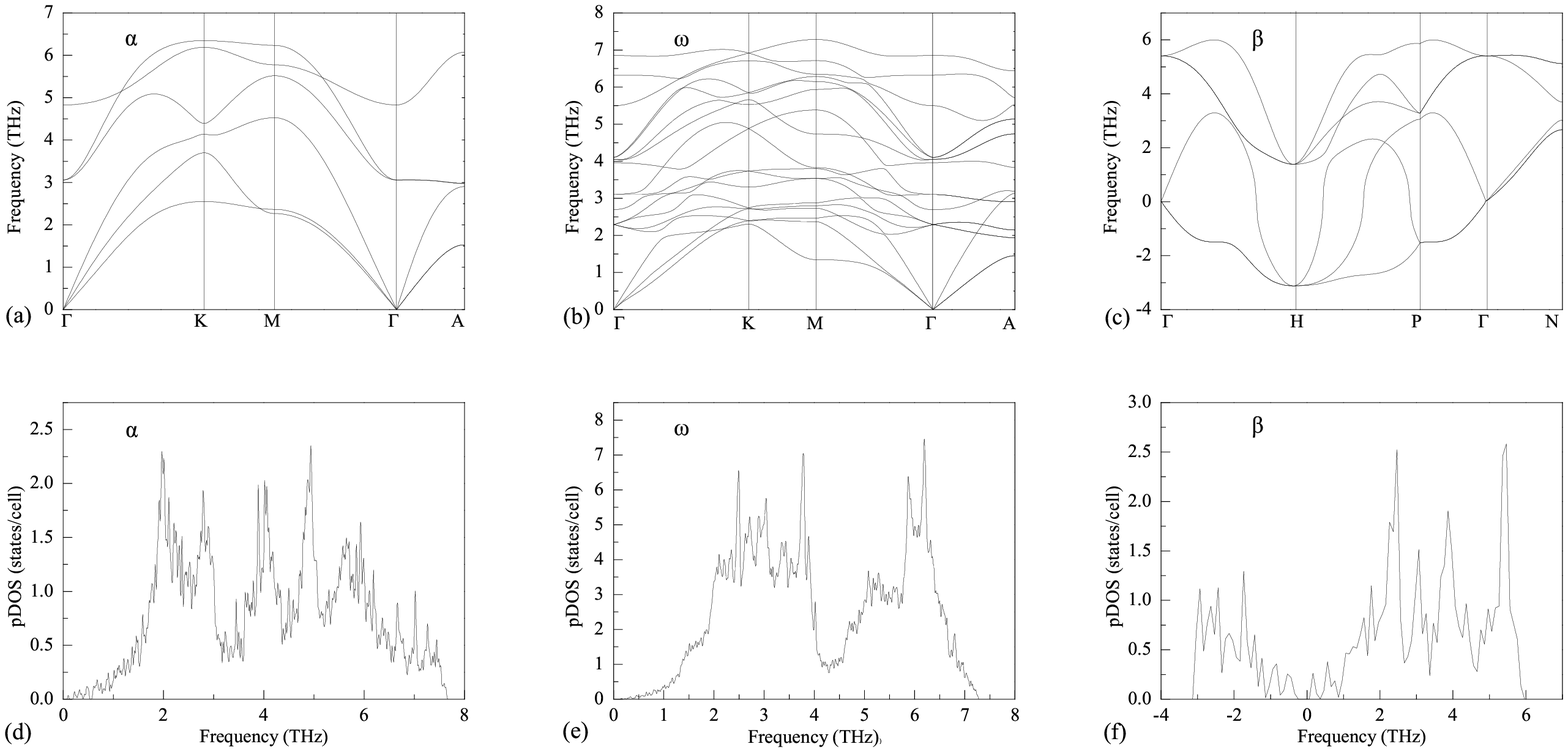}\caption{Calculated Phonon
dispersions (upper panels) and phonon DOS (lower panels) of $\alpha$, $\omega
$, and $\beta$ phases of HfTi alloy. }%
\label{fig:fig6}%
\end{figure}

Elastic constants not only provide valuable information about the
bonding characteristic between adjacent atomic planes and anisotropy
in the bonding, but also can measure the resistance and mechanical
features of crystal to external stress or pressure, which further
describe the stability of crystals against elastic deformation. Our
calculated results of the elastic constants for the three phases of
HfTi alloy are listed in Table I. For comparison, the theoretical
and experimental results of pure Hf and Ti metals are also listed.
Evidently, the $\alpha$ and $\omega$ phases for HfTi alloy as well
as the pure Hf and Ti metals are all mechanically stable at ambient
pressure. However, the $\beta$ phase is unstable for both HfTi alloy
and its archetype metals, since their elastic constants do not
satisfy the mechanical stability criteria of cubic structure
\cite{Nye}. In general, for $\alpha$ and $\omega$ HfTi alloy, our
calculated values of the five independent elastic constants at
ambient pressure lie in the range of corresponding Hf and Ti metals.
With increasing pressure, all the five elastic constants increase
monotonically, among which the $C_{44}$ has a moderate increase,
while other four parameters increase rapidly with applied pressure,
as shown in Fig. 3. We notice that parameter $C_{33}$ of $\omega$
phase is much larger than $C_{11}$, indicating that the bonds
between the nearest neighbors along the $(001)$ plane are much
stronger
than the $(100)$ plane. As for $\beta$-HfTi, the value of $C_{11}%
\mathtt{-}C_{12}$ is negative, coinciding with that of Hf and Ti metals. As
the pressure increases from 0 to 44 GPa, the values of $C_{11}$ and $C_{12}$
increase near linearly, and $C_{11}\mathtt{-}C_{12}$ becomes positive at 18.5
GPa, as shown in Fig. 3.

After obtaining elastic constants at different pressures, the
polycrystalline bulk modulus $B$ and Shear modulus $G$ as functions
of pressure can be evaluated from the Voigt-Reuss-Hill (VRH)
approximation \cite{Voigt, Reuss, Hill}, and the results are
depicted in Fig. 4. For all the three phases of HfTi, the deduced
bulk moduli from VRH approximation at ambient pressure turn out to
be very close to that obtained from the EOS fitting, indicating that
our calculations are consistent and reliable. As the pressure
increases, both bulk modulus and shear modulus increase
monotonically for all the three phases. The increasing rates of bulk
moduli for all the three phases are apparently larger than that of
shear moduli. It is well known that a high (low) ratio of $B/G$ is
responsible for the ductility (brittleness) of polycrystalline
materials. Our calculated values of $B/G$ for $\alpha$ phase
increase from 1.86 to 2.36 under pressure from 0 GPa to 12 GPa, for
$\omega$ phase increase from 1.89 to 3.02 with pressure enhancing
from 0 to 42 GPa, for $\beta$ phase decrease from 16.01 to 9.10 upon
compression from 18.5 GPa to 44.0 GPa. Results show that transition
to $\omega$ phase or $\beta$ phase from $\alpha$ phase will enhance
the ductility, and the $\beta$ phase possess the biggest ductility.
Our calculated $B/G$ for $\alpha$ and $\omega$ phases are 1.62 and
1.66 for Hf, 1.89 and 1.75 for Ti, respectively. We notice that for
the HfTi alloy the ductility is improved with respect to the pure Hf
and Ti metals.

To discuss the pressure induced $s\mathtt{-}d$ electron transfer, we perform
the Mulliken charge population analysis \cite{Mulliken} of the $\beta$ phase.
The variation of the number of electrons on $s$, $p$, $d$ orbitals with
increasing pressure for HfTi alloy and Hf metal are shown in Fig. 5. It is
evident that the $s-d$ electron transfer behavior of HfTi and Hf are
consistent. Upon compression up to 58 GPa, the $d$ electrons of Hf and Ti
atoms in HfTi alloy decrease, while the $s$ and $d$ electrons increases. This
fact is the same as that in Hf metal. The increase of $d$-band occupancy will
stabilize $\beta$ phase of HfTi alloy under pressure.

\subsection{The $P\mathtt{-}T$ phase diagram}

\begin{figure}[ptb]
\includegraphics[width=0.8\textwidth]{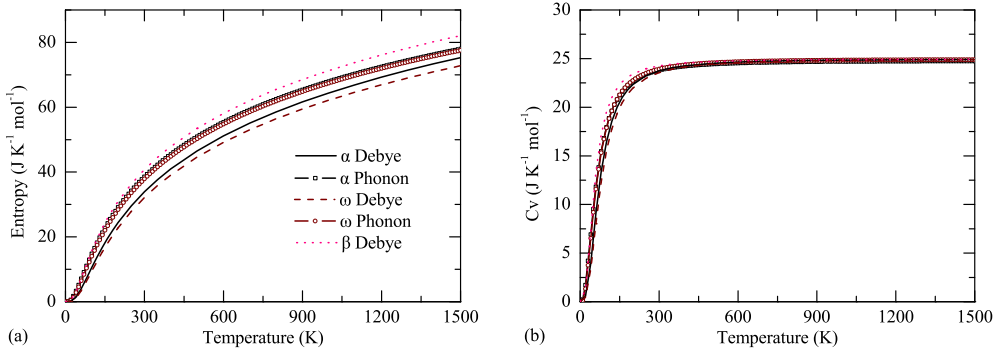}\caption{Temperature dependence of
(a) entropy and (b) specific heat at constant volume for $\alpha$, $\omega$,
and $\beta$ phases of HfTi alloy. }%
\label{fig:fig7}%
\end{figure}

\begin{figure}[ptb]
\includegraphics[width=0.8\textwidth]{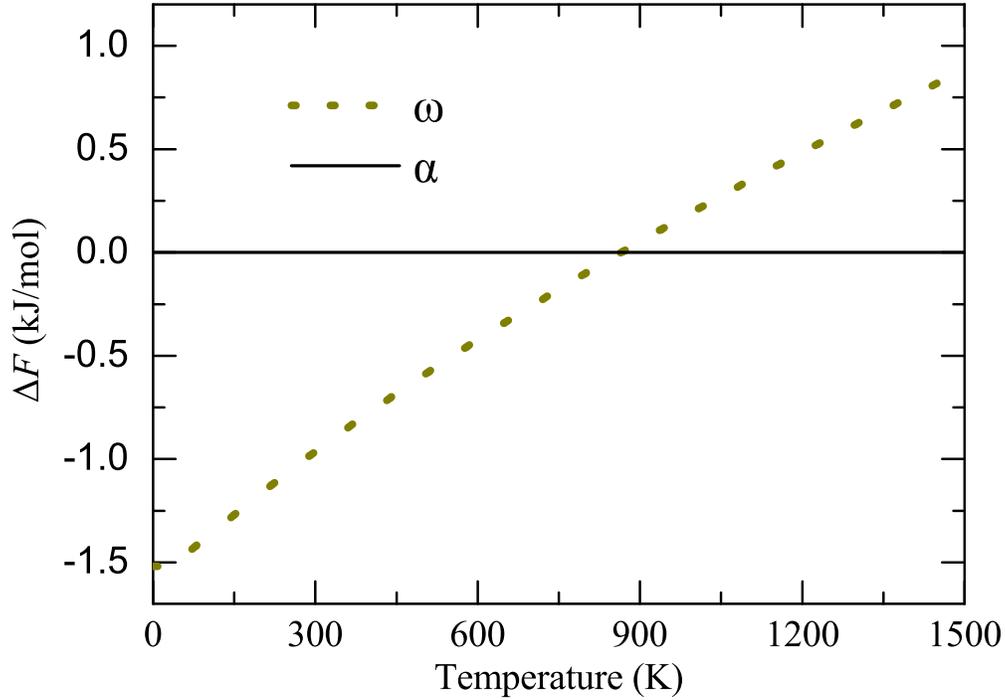}\caption{The Helmholtz free energy
difference ($\Delta F$) of $\omega$ phase HfTi alloy with respect to $\alpha$
as a function of temperature by quasiharmonic approximation at ambient
pressure. }%
\label{fig:fig8}%
\end{figure}

\begin{figure}[ptb]
\includegraphics[width=1.0\textwidth]{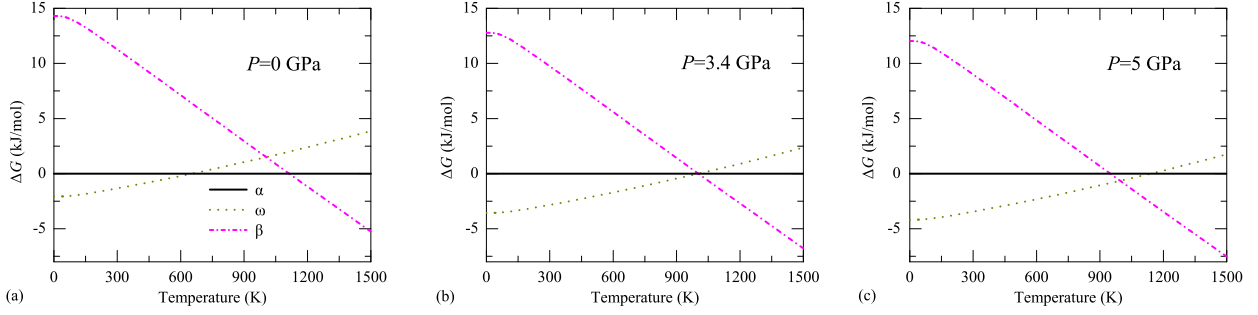}\caption{Temperature dependence of
the Gibbs free energy difference ($\Delta G$) of HfTi phases with respect to
$\alpha$ at different pressures, i.e., 0 GPa, 3.4 GPa, and 5 GPa. }%
\label{fig:fig9}%
\end{figure}

\begin{figure}[ptb]
\includegraphics[width=0.8\textwidth]{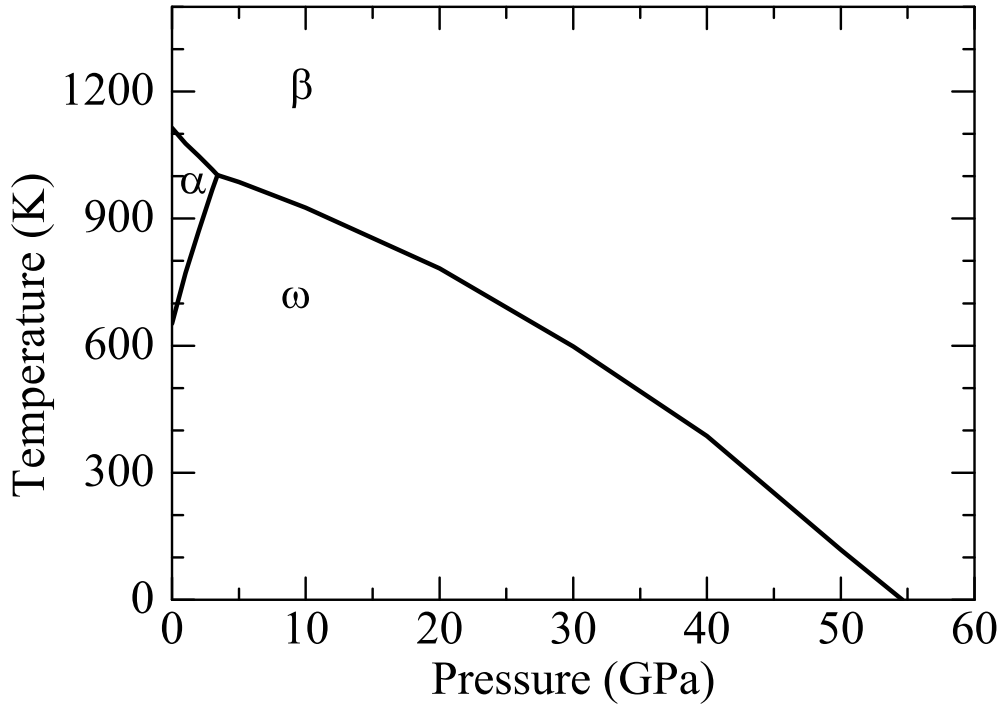}\caption{$P\mathtt{-}T$ phase
diagram of HfTi alloy. The solid lines show our predicted $\alpha
\mathtt{\rightarrow}\omega$, $\omega\mathtt{\rightarrow}\beta$, and
$\alpha\mathtt{\rightarrow}\beta$ transition boundaries by Debye model.}%
\label{fig:fig10}%
\end{figure}The vibration energy of lattice ions can be determined by the
quasiharmonic approximation or the Debye model as specified in Sec II. In
calculating the phonon dispersion curves and the phonon DOS, the
Hellmann-Feynman theorem and the direct method \cite{Parlinski} are employed.
For the BZ integration, the 5$\times$5$\times$5, 3$\times$3$\times$1, and
5$\times$5$\times$5 $k$-point meshes are used for the $\alpha$, $\omega$, and
$\beta$ 3$\times$3$\times$3 supercells, respectively. In Fig. 6 we show the
calculated phonon dispersion curves and phonon DOS of $\alpha$, $\omega$, and
$\beta$ HfTi at ambient pressure. Obviously, the phonon dispersions of
$\alpha$ and $\omega$ phases are stable at ambient pressure, while the $\beta$
phase is unstable. The phonon behavior of $\alpha$-HfTi is similar to that of
metal Hf \cite{Bajpai} and Ti \cite{Mei2}. There are obvious interactions
between acoustic and optical branches in metal Hf and Ti. However, this
character does not appear in their $\alpha$ phase alloy. The interaction
between Hf and Ti atoms in HfTi alloy is weaker than that in its archetype
metals. Due to the low $c/a$ ratio of the $\omega$ phase, the acoustic phonon
branches of Ti metal are stiffer along the $c$ axis than in the basal plane
\cite{Mei2}, and the HfTi alloy shows the same characteristics. As for the
$\beta$ phase, the only stable phonon branch is along [110] direction in HfTi
alloy, which is different from that in metal Ti. The stable nature of phonon
branch along [110] direction for $\beta$-HfTi indicates that the occurrence of
phase transform from $\alpha$ or $\omega$ phase to $\beta$ phase need
considerable external driving force, such as high-temperature or
high-pressure, to break the original phase structure.

The Gibbs free energy, the entropy, and the specific heat at constant volume
($C_{v}$) can be evaluated by both the quasiharmonic approximation and the
Debye model. The entropies of $\alpha$ and $\omega$ phases obtained by
quasiharmonic approximation are both somewhat higher than those obtained by
Debye model, as shown in Fig. 7(a). The calculated $C_{v}$ with quasiharmonic
approximation is almost identical to that obtained by Debye model for $\alpha$
phase, while the difference between these two schemes for $\omega$ phase is
slightly larger [Fig. 7(b)]. At ambient pressure the Gibbs free energy is
equal to the Helmholtz free energy. As shown in Fig. 8, we calculate the
Helmholtz free energy of $\omega$ phase with respect to the $\alpha$ phase as
a function of temperature by quasiharmonic approximation. Note that the free
energy calculations include the zero-point energy (2.483 kJ/mol for $\omega$
phase and 2.444 kJ/mol for $\alpha$ phase). A $\omega$ to $\alpha$ phase
transition temperature of 865 K can be obtained. Since the $\beta$ phase is
thermodynamically unstable, we are unable to predict the lattice vibrational
energy of the $\beta$ phase by phonon DOS. Alternatively, we calculate the
Gibbs free energy by the Debye model. Figure 9 shows the calculated Gibbs free
energy of $\omega$ and $\beta$ phases with respect to the $\alpha$ phase as a
function of temperature at different pressures, \textit{i}.\textit{e}., 0 GPa,
3.4 GPa, and 5 GPa. At zero pressure the $\omega$ phase has the lowest Gibbs
energy within the temperature range 0 K$<$$T$$<$653 K, and this transition
temperature is lower than that deduced from quasiharmonic approximation by
$\sim$200 K. From 653 K to 1112 K, the $\alpha$ phase is preferred, and this
result is consistent with the experimental temperature range of 500-1165 K
\cite{Bittermann}. When the temperature is further increased to be beyond 1112
K, the $\beta$ phase becomes stable. At $P$=3.4 GPa, the three phases have the
same Gibbs energy at 1003 K, corresponding to a triple point in the
$P\mathtt{-}T$ phase diagram. Above 3.4 GPa, the $\omega$ phase transits to
$\beta$ phase without formation of $\alpha$ phase [Fig. 9(c)]. Finally, in
Fig. 10 we depict the $P\mathtt{-}T$ phase diagram of HfTi alloy calculated by
Debye model. Remarkably, our calculated $\alpha\mathtt{\rightarrow}\beta$
transition temperature at ambient pressure (1112 K) is very close to the
experimental measurement (1203$\pm$31 K) \cite{Ruda}. The triple point is
predicted to be (3.4 GPa, 1003 K), which needs experimental test to identify
in the future.

\section{CONCLUSIONS}

In summary, the structural phase transition, pressure-dependent elasticity
behavior, and phonon spectra of HfTi alloy have been theoretically studied.
The obtained ground-state structural parameters of $\alpha$ and $\omega$
phases of HfTi alloy and its archetype metals are consistent well with
accessible experimental data. The calculated elastic constants indicate that
the $\alpha$ and $\omega$ phases are mechanically stable at ambient pressure,
while the $\beta$ phase is unstable. The values of elastic constants for all
the three phases of HfTi alloy are between Hf and Ti metals. The
$\alpha\mathtt{\rightarrow}\beta$ alloy phase transition pressure is predicted
to be 54.3 GPa, which is close to that for Hf metal. Under compression,
elastic constants, bulk modulus $B$, and shear modulus $G$ increase almost
linearly for all the three phases. The $\beta$ phase become mechanically
stable at 18.5 GPa. The Mullikey charge population analysis shows that the $p$
electrons of Hf and Ti in $\beta$ HfTi alloy transfer to corresponding $d$ and
$s$ orbitals upon compression, which strengthens the stability of $\beta$
phase under pressure, and this character is similar to that in $\beta$-Hf
metal. The nature of stability for $\alpha$ and $\omega$ phases at ambient
pressure has also been observed from phonon dispersions. As for the $\beta$
phase, the only stable phonon branch is along the [110] direction, which is
different from that in metal Ti. The lattice vibrational energy was calculated
based on quasiharmonic approximation from both the phonon DOS and Debye model.
As a consequence, thermodynamic properties of Gibbs free energy, entropy, and
specific heat at constant volume of $\alpha$- and $\omega$-HfTi have been
theoretically obtained. The transition temperature of $\omega
\mathtt{\rightarrow}\alpha$ at ambient pressure is 865 K by phonon and 653 K
by Debye model, respectively. Finally, based on the Gibbs free energy
evaluated from Debye model as functions of pressure and temperature, the
$P\mathtt{-}T$ phase diagram has been depicted. Remarkably, our predicted
$\alpha\rightarrow\beta$ phase transition temperature of 1112 K coincides well
with the attainable experimental report.

\begin{acknowledgments}
This work was supported by NSFC under Grant No. 51071032, and by Foundations
for Development of Science and Technology of China Academy of Engineering
Physics under Grants No. 2011A0301016.
\end{acknowledgments}

\end{document}